\documentstyle[preprint,aps]{revtex}
\tightenlines
\begin{document}
\draft

\title{Theory of hopping magnetoresistance induced by Zeeman splitting}

\author{Penny Clarke,$^{(1)}$ L. I. Glazman,$^{(1)}$ and K. A.
Matveev$^{(2)}$}

\address{
$^{(1)}$Theoretical Physics Institute and Department of Physics,
University of Minnesota, Minneapolis, Minnesota 55455\\
$^{(2)}$Massachusetts Institute of Technology, 12-105, Cambridge,
Massachusetts 02139}
\maketitle
\begin{abstract}
We present a study of hopping conductivity for a system of sites which
can be occupied by more than one electron. At a moderate on-site
Coulomb repulsion, the coexistence of sites with occupation numbers 0,
1, and 2 results in an exponential dependence of the Mott conductivity
upon Zeeman splitting $\mu_BH$.  We show that the conductivity behaves as
$\ln\sigma= (T/T_0)^{1/4}F(x)$, where $F$ is a universal scaling
function of $x=\mu_BH/T(T_0/T)^{1/4}$. We find $F(x)$ analytically at
weak fields, $x \ll 1$, using a perturbative approach.  Above some
threshold $x_{\rm th}$, the function $F(x)$ attains a constant value,
which is also found analytically.  The full shape of the scaling
function is determined numerically, from a simulation of the
corresponding ``two color'' dimensionless percolation problem.  In
addition, we develop an approximate method which enables us to solve
this percolation problem analytically at any magnetic field.  This
method gives a satisfactory extrapolation of the function $F(x)$
between its two limiting forms.
\end{abstract}

\section{Introduction}
Low-temperature conductivity in a disordered semiconductor is controlled
by phonon-assisted electron hops between localized states. At sufficiently
low temperatures, only those sites which have energy levels close to the Fermi
level participate in the hopping transport. This defines the
Mott variable range hopping (VRH) regime\cite{Mott}. Mott
conductivity depends exponentially on temperature, $\sigma (T)\propto
\exp[-(T_0/T)^{1/(d+1)}]$, where $T_0=\beta_d/ga^d$ is the characteristic Mott
temperature, $g$ is the density of localized states at the Fermi level, $a$ is
the localization radius for a single site, $d$ is the dimensionality of the
sample, and the numerical factor $\beta_d$ is determined by percolation
theory\cite{ShklEfr}. In the more standard case of lightly doped
semiconductors, the strip of localized states in energy space is relatively
narrow, and each site can accommodate at most one electron. Under these
conditions, the spin degree of freedom of the hopping electrons has no effect
upon the exponential  factor of the hopping conductivity.  The application of
a magnetic  field $H$ results in the modification of this exponential
factor due solely to the  orbital effect of the field\cite{ShklEfr}.

Kamimura {\em et al.}\cite {Kamimura} were the first to recognize that the
spin degree of freedom plays a significant role in the magnetoresistance if a
certain fraction of the sites can accommodate more than one electron.  Double
occupancy is possible if the on-site Coulomb repulsion $U$ between the
electrons is smaller than the width of the distribution function of the
energies of the localized sites. In this case there are two types of sites
which contribute to hopping transport. Sites of the first type, which we will
call type $A$, have energies $\epsilon$ close to $\mu$. Sites of type $B$ have
one electron at a deep level with $\epsilon\sim \mu-U$, so that the energy for
the second electron is close to $\mu$. The sites which are neither of type $A$
nor of type $B$ have energy levels which are too far from the Fermi level to
contribute to transport. At zero magnetic field, the probability for two
electrons on two singly-occupied sites to have opposite spins equals $1/2$.
Therefore, hops between $A$ and $B$ sites can occur.  In the strong field
limit however, all spins are polarized, and $A\leftrightarrow B$ hops are
completely suppressed\cite{Kamimura}, assuming that two electrons occupying
the same site form a singlet state at all relevant magnetic
fields\cite{singlet}. Thus, at sufficiently strong magnetic fields the
characteristic Mott temperature is determined by the larger of the two
densities of states $g_A$ and $g_B$ rather than by the net density of states
$g=g_A+g_B$. The increase in $T_0$ due to the field induced suppression of
$A\leftrightarrow B$ hops leads to a giant positive magnetoresistance.

In this paper we present a detailed study of the hopping magnetoresistance
induced by Zeeman splitting.  We show that the criterion for the strong
magnetic field limit described above corresponds to a {\em finite} threshold
value $H_{\rm th}$.  Below this threshold value, the conventional Mott
exponent, $(T_0/T)^{1/(d+1)}$, is modified by a factor $F(x)$,
\begin{equation}
\sigma \propto \exp\left\{-\left(\frac{T_0}{T}\right)^{1/(d+1)}F(x)\right\},
\label{newmott}
\end{equation}
which is a {\em universal} function of a single scaling parameter
\begin{equation}
x = \frac{\mu_B H}{T (T_0/T)^{1/(d+1)}}.
\label{scaling}
\end{equation}
The universality of the Zeeman splitting induced magnetoresistance is a
key result of this paper.  Its significance is illuminated by noting that it
reduces
the calculation of the magnetoresistance to the determination of a universal
function $F(x)$ of a single dimensionless parameter $x$, as opposed to
previous solutions\cite{Kamimura} for which the magnetoresistance was
calculated as a function of two variables, $T$ and $H$.

Above a certain threshold $x\geq x_{\rm th}$, the universal function
$F(x)$ attains a constant value which we determine analytically.  In
addition to the saturation value, we find the $x \ll 1$ asymptote of
$F(x)$ analytically using a perturbative approach.  In an attempt to
obtain an analytically determined fitting curve for the universal
scaling function $F(x)$, we extend the invariant technique
\cite{ShklEfr} to the case of the ``two-color'' percolation problem.
To determine the accuracy of these two analytical techniques, we find
the scaling function $F(x)$ numerically by simulations of the
corresponding ``two color'' dimensionless percolation problem and
compare it with the functions obtained analytically.

Some of the results of this study have been reported in a short
communication\cite{short}. Another mechanism of giant
magnetoresistance due to two types of localized states was considered
in Ref.\cite{Gogolin}.

\section{Applicability of the model}
\label{Apply}
Zeeman splitting can make the dominant contribution to the magnetoresistance
in a number of important cases, which include those of undoped amorphous
silicon\cite{Kamimura,Ephron} and moderately disordered two-dimensional
electron systems. In the latter case, the minima of the random potential may
serve as sites accommodating several electrons.  The orbital
effect can be avoided altogether if the field $H$ is applied parallel to the
plane of the two-dimensional system.  In amorphous silicon electrons are
localized at ``dangling bonds''\cite{Elliott}.  For the case of bulk
$\alpha$-Si, the orbital effects of the  magnetic  field cannot be
eliminated, and Zeeman splitting makes the dominant  contribution to the
magnetoresistance only in a certain range of  temperatures.  To determine this
range, we consider the tunneling of an  electron a distance $L$ through a
barrier consisting of the superposition of  an ``intrinsic'', $V_0$, and a
``magnetic'', $m \omega_c^2 x^2 /2$, barrier  ($\omega_c=eH/mc$ is the
cyclotron
frequency of the field $H$).  The action  for the electron subbarrier motion is
\begin{equation}
\frac{S}{\hbar}= \frac{1}{\hbar} \int \sqrt{2m \left( V_0 +
	\frac{1}{2}m \omega_c^2 x^2 \right)} dx \approx \frac{L}{a} +
	\frac{a L^3}{ 6 \lambda^4}.
\label{action}
\end{equation}
The localization radius of the electron in the absence of a magnetic field is
related to the strength of the ``intrinsic'' barrier $V_0$ as
$a=\hbar/\sqrt {2mV_0}$, and $\lambda=\sqrt{c \hbar/ eH}$ is the
magnetic length. The second term in Eq. (\ref{action}) represents the
correction to the subbarrier action due to the presence of a magnetic field.
{}From Eq. (\ref{action}), we see that the orbital effect of the magnetic field
is negligible at
\begin{equation}
\frac{a L^3}{6 \lambda^4} \ll 1.
\label{orbital}
\end{equation}
The characteristic tunneling distance $L$ in the VRH regime is of course
determined by the temperature
\begin{equation}
L \approx a \left( \frac{T_0}{T} \right)^{1/(d+1)}.
\label{temp}
\end{equation}
Eqs.~(\ref{orbital}) and (\ref{temp}) determine the upper limit
of the magnetic field. Requiring the Zeeman splitting to be strong limits the
magnetic field from below:
\begin{equation}
T \ll \mu_B H \ll \frac{ \sqrt6 \hbar^2}{2ma^2}\left( \frac{T}{T_0}\right)^
	{3/{2(d+1)}}.
\end{equation}
This relation places the restriction
\begin{equation}
T<T_0 \left( \frac{\hbar^2} {m a^2 T_0} \right)^{2(d+1)/(2d-1)}.
\label{restriction}
\end{equation}
upon the range of temperatures at which Zeeman splitting makes the
dominant contribution to the magnetoresistance.

\section{Derivation of the Connectivity Conditions}

Mott's arguments enable one to find the temperature dependence of the
exponential factor in VRH conductivity. However, it is necessary to use
percolation theory to determine the numerical factor $\beta_d$ in the
exponent\cite{ShklEfr}.  Similarly, to study quantitatively the effect of
Zeeman splitting upon the conductivity one has to reformulate the problem in
terms of percolation theory. Because two different site types are now
involved, the percolation network which determines the hopping conductivity
consists of three types of links: $AA$, $BB$, and $AB$. In order to find the
VRH conductivity, we must first determine the conductances $G_{AA}$, $G_{BB}$,
and $G_{AB}$ of the three types of elementary links.  To begin with, we
introduce the probabilities $P_i$ of having the site $i$ occupied by zero
($0$), one ($\uparrow$ or $\downarrow$), or two ($\uparrow \downarrow$)
electrons:
\begin{eqnarray}
P_i(0)=Z^{-1},
\label{P0}\\
\quad P_i(\uparrow)=Z^{-1} \exp[-(\epsilon_i + \mu_B H)/T],
\label{P1}\\
\quad P_i(\downarrow)=Z^{-1}\exp[-(\epsilon_i - \mu_B H)/T],
\label{P-1}\\
\quad P_i(\uparrow \downarrow)=Z_{-1} \exp[-(2 \epsilon_i + U)/T].
\label{P2}
\end{eqnarray}
Here
\begin{equation}
Z=1 + \exp[-(\epsilon_i +\mu_B H)] + \exp[-(\epsilon_i-\mu_B H)]
+ \exp[-(2 \epsilon_i + U)/T].
\label{partition}
\end{equation}
Neglecting the pre-exponential factors (which include, e.g., the
deformation potential constant), we can express the three elementary
link conductances as products of phonon and electron occupation
numbers.
\begin{eqnarray}
G_{AA} \propto [P_1(\uparrow) + P_1(\downarrow)]P_2(0)
N(\epsilon_2-\epsilon_1) \exp(-2R/a),
\label{GAA}\\
G_{BB} \propto P_1(\uparrow \downarrow)[P_2(\uparrow)+P_2(\downarrow)]
N(\epsilon_2-\epsilon_1) \exp(-2R/a),
\label{GBB}\\
G_{AB} \propto [P_1(\uparrow)P_2(\downarrow)+P_1(\downarrow)P_2(\uparrow)]
N(\epsilon_2+U -\epsilon_1) \exp(-2R/a).
\label{GAB}
\end{eqnarray}
Here $R$ is the distance between two sites and $N$ is the Bose distribution
function.

The standard approach to the formulation of the percolation problem requires
the exponential representation of the elementary link conductance, $G \propto
e^{- \xi}$.  In this colored percolation problem, there are three different
exponents $\xi_{AA}$, $\xi_{BB}$, and $\xi_{AB}$ which are extracted from
Eqs.~(\ref{GAA})--(\ref{GAB}).  Assuming that the on-site Coulomb interaction
is very strong $U \gg \mu_B H \gg T$, we find
\begin{eqnarray}
\xi_{AA} &=& \frac{|\epsilon_2 - \epsilon_1| + |\epsilon_1 - \mu_B H|
              + |\epsilon_2 - \mu_B H|}{2T} + \frac{2R}{a},
\label{xiAA}\\
\xi_{BB} &=& \frac{|\epsilon_2 - \epsilon_1| + |\epsilon_1 + U + \mu_B H|
              + |\epsilon_2 + U + \mu_B H|}{2T}
\nonumber\\
          &&+ \frac{2R}{a},
\label{xiBB}\\
\xi_{AB} &=& \frac{|\epsilon_2  + U - \epsilon_1| + |\epsilon_1 - \mu_B H|
     + |\epsilon_2 + U + \mu_B H|}{2T}
\nonumber\\
          &&+\frac{\mu_B H}{T}+\frac{2R}{a}.
\label{xiAB}
\end{eqnarray}
The energies $\epsilon_1$ and $\epsilon_2$ in Eq. (\ref{xiAB}) correspond to
the energy levels of sites of type $A$ and $B$ respectively.  The Fermi level
depends upon magnetic field due to transitions from doubly to singly occupied
sites, $\delta\mu=\mu_{B}H(g_B-g_{A})/(g_{A}+g_{B})$.  However, this has no
effect upon the resistance, provided that $g_{A}$ and $g_{B}$ are energy
independent for energies on the order of $\mu_{B}H$.  Unlike the case of
semiconductors doped by shallow impurities\cite{ShklEfr}, in amorphous silicon
the densities of states are independent of energy on this scale, since they
vary with energies comparable to the onsite Coulomb interaction $U \approx
100$ meV\cite{Ephron2}.  Therefore, we can safely make the assumption of
constant densities of states in the vicinity of the Fermi level.

Instead of the energies $\epsilon_i$ it is convenient to introduce a new set
of variables $\varepsilon_i$, chosen in such a way that their values are close
to zero for sites participating in electron transport:
\begin{equation}
\begin{cases}
{\varepsilon = \epsilon - \mu_B H  & for $A$-sites,
\cr \varepsilon = \epsilon
+ U + \mu_B H & for $B$-sites.}
\end{cases}
\label{varepsilon}
\end{equation}
We further define
\begin{equation}
\varepsilon_{12}(H)=\frac{|\varepsilon_2-\varepsilon_1-2\mu_BH| +
	|\varepsilon_1| +|\varepsilon_2|}{2} + \mu_BH.
\label{var12}
\end{equation}
The new variables simplify the dimensionless exponents
(\ref{xiAA})--(\ref{xiAB}) to the forms:
\begin{equation}
\xi_{AA} =\xi_{BB} = \frac{\varepsilon_{12}(0)}{T} + \frac{2R}{a},
\label{xiAABB}
\end{equation}
\begin{equation}
\xi_{AB} \!=\! \frac{\varepsilon_{12}(H)}{T}+\frac{2R}{a}.
\label{xiABAB}
\end{equation}
In the absence of a magnetic field, $\xi_{AA} =\xi_{BB}= \xi_{AB}$, and we
return to the standard percolation problem for VRH conductivity. At $H=0$
only the net density of states $g = g_{A} + g_{B}$ at the Fermi level is
relevant, and the ratio of the densities of states $g_{A}/g_{B}$
does not affect the conductivity.

\section{Magnetoresistance in the Limits of Weak and Strong Fields}
\subsection{Strong Field Limit}
As follows from Eq.~(\ref{xiABAB}), the exponent $\xi_{AB}$ can not be
smaller than $2\mu_B H/T$. Thus at sufficiently strong magnetic fields the
transitions between sites $A$ and $B$ cannot occur\cite{Kamimura}. In this
limit, the conductivity is determined by two parallel
percolation networks, one of which consists only of type $A$ sites and
the other of only type $B$ sites. Therefore the
conductivity is independent of $H$ and satisfies Mott's law with
the density of states $\tilde g = \max\{g_A, g_B\}$. This picture is valid
above a certain threshold for the magnetic field:
\begin{equation}
 H>H_{\rm th} \equiv \frac{T}{2\mu_B} \tilde \xi_c,
\label{strongfield}
\end{equation}
when all possible $\xi_{AB}$  exceed the critical exponent $\tilde \xi_c$
determined by the percolation theory solution of the VRH
problem \cite{ShklEfr},
\begin{equation}
\tilde \xi_c = \left[\frac{(g_A+g_B) T_0}{\tilde g T}\right]^{1/(d+1)},
\quad T_0=\frac{\beta_d}{(g_A+g_B)a^d}.\nonumber
\end{equation}

\subsection{Weak Field Limit}
\label{weak}
The opposite limit of weak fields also allows for analytical consideration
by means of a perturbative approach applied to the percolation problem with
the density of states $g=g_A+g_B$ and the percolation threshold $\xi_c^0 =
(T_0/T)^{1/(d+1)}$. At $H\ll H_{\rm th}$ possible field induced variations of
$\xi_{AB}$ are small, $\Delta \xi_{AB} \ll \xi_c^0$. This enables us to
use the perturbative approach proposed in
\cite{ShklEfr}. According to Ref.\cite{ShklEfr}, one can find the small shift
of the percolation threshold $\xi_c$ as an average increment of $\xi$ caused
by the small perturbation, $\Delta \xi_c = \langle \Delta \xi \rangle$. To
calculate the average $\langle
\Delta \xi \rangle$ over the statistical ensemble of sites, one needs to
find the correction $\Delta \xi$ to the exponent for each link due to a small
magnetic field.  Clearly, $\Delta \xi_{AA}= \Delta \xi_{BB} =0$, whereas
$\Delta \xi_{AB}= 2\mu_B H/T$ if $\varepsilon_1 > \varepsilon_2$, and $\Delta
\xi_{AB}= 0$ otherwise, see Eqs.~(\ref{xiAABB}) and (\ref{xiABAB}). Taking
into  account the fraction of $AB$ links with $\varepsilon_1 > \varepsilon_2$
in the percolation cluster $g_A g_B /(g_A +g_B)^2$, we find
\begin{equation}
\Delta \xi_c = \langle \Delta \xi \rangle = \frac{2g_A g_B}{(g_A+g_B)^2}
               \frac{\mu_B H}{T}.
\label{deltaxic}
\end{equation}
The dependence of the hopping conductivity upon the critical exponent
$\xi_c$ is  $\sigma \sim e^{-\xi_c}$.
Therefore Eq.~(\ref{deltaxic}) enables us to find the dependence of the
conductivity on the magnetic field:
\begin{equation}
\ln{\frac{\sigma (T, H)}{\sigma (T, 0)}}=-\frac{2g_A g_B}{(g_A+g_B)^2}
                \frac{\mu_B H}{T}.
\label{sigmaTH}
\end{equation}
This dependence is sensitive to the relative densities of $A$- and
$B$-sites. As expected, there is no field dependence at $g_A=0$ or
$g_B=0$.

The region of validity of (\ref{sigmaTH}) is determined by the
applicability of the perturbative approach, i.e., by the requirement
$\Delta \xi_{AB} \ll \xi_c^0$. In terms of the magnetic field strength, this
condition reads $\mu_B H \ll \mu_B H_{\rm th} \sim T \xi_c^0$.
Since $\xi_c^0 \gg 1$, the
latter condition does not contradict our initial assumption, that
$\mu_B H\gg T$. More precise limits of applicability of the perturbative
calculation become apparent from the comparison of Eq. (\ref{sigmaTH}) with
the results of numerical simulation which we present later in this paper.

\section{Scaling Analysis}
\subsection{Scaling Conjecture}

It is worth noting that the dimensionless parameter of the
perturbation theory is
\begin{equation}
x = \frac{\mu_B H }{T\xi_c^0} \equiv \frac{\mu_B H}{T (T_0/T)^{1/(d+1)}}.
\label{x}
\end{equation}
The threshold field $H_{\rm th}$ corresponds to the universal (i.e.,
temperature independent) value of this parameter,
\[
x_{\rm th} = \frac12 \left(\frac{g_A +g_B}{\tilde g}\right)^{1/(d+1)}.
\]
The existence of the dimensionless parameter $x$  allowed us to make the
conjecture (\ref{newmott}), which corresponds to the following scaling
behavior of the percolation threshold: $\xi_c (T, H)=\xi_c^0 F(x)$.  This
scaling function $F(x)$ determines the conductivity at finite magnetic fields
\begin{equation}
\sigma \propto \exp \left\{  - \left( \frac{T_0}{T} \right)^{1/(d+1)}F
\left( \frac{\mu_B H}{T(T_0/T)^{1/(d+1)}} \right) \right\}.
\label{scaling1}
\end{equation}
{}From the cases of low
and high fields discussed above we already know the limiting behavior of the
function $F(x)$. It has linear expansion at small $x$ and reaches a constant
value at large $x$:
\begin{equation}
F(x) = \begin{cases}
{1+\frac{2g_A g_B}{(g_A+g_B)^2}x & at $x \ll 1$,  \cr
\left(\frac{g_A +g_B}{\tilde g}\right)^{1/(d+1)}& at $x \geq x_{\rm th}$.}
\end{cases}
\label{asympt}
\end{equation}
\subsection{Proof of the Scaling Conjecture}
We will now prove our conjecture (\ref{newmott}), (\ref{scaling}) by reducing
the initial problem of hopping magnetoresistance to a dimensionless
percolation problem.

We start from the conventional percolation approach to the hopping
conductivity and introduce a positive variable $\xi$. At given $\xi$, all
links are cut except for those with conductance
$G>e^{-\xi}$. At small $\xi$ percolation does not occur,
 but at some particular value $\xi = \xi_c$ the network starts
to percolate. This threshold value determines the conductivity $\sigma
\sim e^{-\xi_c}$. Our goal is to find the dependence of $\xi_c$ on $T$ and
$H$. We generalize the approach of Ref.~\cite{ShklEfr} and introduce
a set of dimensionless variables $\Delta$, $\rho$, and $\chi$ defined by
the following relations:
\begin{equation}
\varepsilon = T \xi \Delta, \quad
R = \frac12 a \xi \rho, \quad
\mu_B H = T \xi \chi.
\label{defchi}
\end{equation}
In these variables the connectivity condition for a link is:
\begin{equation}
\begin{array}{ll}
\Delta_{12}(0)+\rho < 1 & \mbox{for $AA$ \& $BB$ links,}\\
\Delta_{12}(\chi)+\rho < 1 & \mbox{for $AB$ links}
\end{array}
\label{connectivity}
\end{equation}
where
\[
\Delta_{12}(\chi)=\frac{|\Delta_1|+|\Delta_2|+|\Delta_1-\Delta_2 -2\chi|}
                       {2} +\chi.
\]
Clearly, only sites with the dimensionless energies $|\Delta| < 1$ can
be connected. The dimensionless concentration of these sites in
$\rho$-space is
\begin{equation}
n= \frac{1}{2^{d-1}}(g_A +g_B) a^d T \xi^{d+1}.
\label{concentration}
\end{equation}
We are now in a position to formulate the dimensionless percolation
problem associated with the hopping magnetoresistance problem.
Consider a random distribution of points with concentration $n$ in
a $d$-dimensional $\rho$-space. Each point is characterized by its
type ($A$ or $B$), radius-vector {\boldmath $\rho_i$}, and energy $\Delta_i$.
The latter is distributed uniformly over the interval $(-1, 1)$. Two points
form a link if the condition (\ref{connectivity}) is satisfied. The problem is
characterized by two dimensionless parameters: the ratio $\gamma$ of
concentrations of $A$- and $B$-points, and dimensionless magnetic field
$\chi$. At $\gamma = g_A/g_B$ this dimensionless problem is equivalent to the
original percolation problem for the hopping magnetoresistance. To reach the
percolation threshold, we increase the total dimensionless concentration $n$
(holding $\gamma$ constant) until the critical value $n_c (\chi)$ is reached.
Once $n_c(\chi)$ is found, we can determine the threshold $\xi_c$ for the
original problem from the relation (\ref{concentration}), which can be
rewritten as
\begin{equation}
n_c(\chi)/n_c(0) = (\xi_c/\xi_c^0)^{d+1}.
\end{equation}
Rewriting $\chi$ as
defined by Eq.~(\ref{defchi}) in terms of the parameter $x$ (see Eq.
(\ref{x})) we find that $\xi_c$ must be a solution of the following equation:
\begin{equation}
\frac{n_c(x\xi_c^0/\xi_c)}{n_c(0)} =
				\left(\frac{\xi_c}{\xi_c^0}\right)^{d+1}.
\end{equation}
One can easily show that this equation has exactly one solution. Then it
obviously has the form
\begin{equation}
\xi_c=\xi_c^0 F(x).
\end{equation}
This proves our scaling conjecture (\ref{newmott}), (\ref{scaling}).

\section{Simulation}
\label{simulation}
As was previously mentioned, the accuracy of the perturbative approach
used at small $x$, see Eq. (\ref{asympt}) can only be determined by
comparison  with simulation data.  In addition, there is an intermediate
range of  field strengths within which $F(x)$ must be numerically determined,
as the analytical theory does not extend to this range.  To perform these
tasks, we developed a code which closely resembles that of Skal {\it et
al.}~\cite{skal}.  We will describe the method as applied in two dimensions.
The simulation  consists of the following: $N$ sites  are randomly thrown in
an $l\times l$ box (which is large enough for the system to be well below the
percolation threshold for the given value of $N$). Each of these sites is
randomly assigned a site type, $A$ or $B$, and an  energy $\Delta$.  Two
sites are connected if their parameters satisfy the condition
(\ref{connectivity}). Percolation is said to occur when two strips (one at
each end of the box), of width equal to the average distance between
sites, are connected. We start with $l \gg l_c(\chi)$, and make $l$ smaller
until this connection occurs. This determines the percolation threshold
$l_c(\chi)$. The critical site concentration is $n_c(\chi)=N/l_c^2(\chi)$.

We ran the simulation at $\gamma=1$ and at $\gamma=1/2$ on $3600$ sites and
averaged each point over $100$ runs. One can see from the data shown in
Fig.~1, that $F(x)$ is an increasing monotonic function which attains a
limiting value for $x\geq x_{\rm th}$. At $\gamma=1$ the limiting value of
$F(x)$ is systematically suppressed below the exact result given by Eq.
(\ref{asympt}).  This is an artifact of the simulation which occurs because
$\gamma=1$ corresponds to the same number of $A$ and $B$ sites. For $x \ge
x_{\rm th}$ at $\gamma=1$, there are two parallel networks which can
percolate. Thus for $x\ge x_{\rm th}$, the simulation is in effect being run
twice, once on the $A$-network and once on the $B$-network, with the smaller
value of the two resulting concentrations selected.
Clearly, this finite size effect does not exist at any other value of
$\gamma$. The simulation data of Fig. 1 used in conjunction with the scaling
functional form (\ref{newmott}) yields the magnetoresistance for $\gamma=1$
and $\gamma=1/2$ at arbitrary fields.

In order to determine the range of our small $x$ asymptotic form, we replotted
the data as critical site concentration $n_c$ vs. $\chi$ and found that
$n_c(\chi)$  begins to deviate from being linear in $\chi$ for $\chi \agt
0.20$.  Weighted  linear fits to the data yielded the zero field values
$n_c(0)=7.064 \pm 0.023$  at $\gamma=1$ and $n_c(0)=7.036 \pm 0.031$ at
$\gamma=1/2$ both of which are  in good agreement with the accepted value of
$6.9 \pm 0.4$\cite{skal}. To test the accuracy of our
weak field perturbative approach, we performed a weighted linear fit to the
data taken at and below $\chi=0.20$.  As one can
see from the insets in Fig. 1(a) and 1(b), the fits to the simulation data
yielded slopes and intercepts which were in excellent agreement with those
obtained perturbatively at both $\gamma=1$ and $\gamma=1/2$.

\section{Approximate Invariant for the Percolation Problem}

Ideally we would like to know the scaling function $F(x)$ for any
value of $\gamma$.  However, the simulations used to determine $F(x)$
for each value of $\gamma$ require a large amount of computer time.
Thus we only found $F(x)$ for two values of gamma: $\gamma=1$ and
$\gamma=1/2$.  The development of an analytical method which yields an
approximate form of the scaling function $F(x)$ is therefore highly
desirable.  The invariant method developed by Shklovskii and
Efros\cite{ShklEfr} gives just such a simple approximate solution of
the VRH problem.  Unfortunately, this approach cannot be adapted to
the case of the two color percolation problem. In this paper we
suggest an alternative invariant.  This approach (i) gives a better
estimate of the constant $\beta_d$ for the standard percolation
problem, and (ii) can be generalized to the two-color case. As it
reduces the problem of determining the universal scaling function
$F(x)$ to the solution of a simple integral equation, we have used it
to obtain an analytically determined fitting curve for the universal
scaling function $F(x)$.

Recently the same invariant conjecture was independently proposed by
Ioselevich\cite{Isol}.

\subsection{Invariant approach to the standard VRH Problem}
\label{VRH_invariant}

We will first formulate our invariant conjecture as applied to the
standard (single-color, $g_B=0$) VRH problem. At the percolation threshold
($\xi=\xi_c$), the probability density of  finding a defect state
with energy $\varepsilon_2$ linked to one with energy $\varepsilon_1$
is
\begin{eqnarray}
G_{AA}(\varepsilon_1,\varepsilon_2)&=&\int \theta(\xi_c-
			\xi_{AA}(\varepsilon_1,\varepsilon_2,r))d^d r
\nonumber\\
&=&v_d g_A \left(\frac{a\xi_c}{2}\right)^d
	\left(1-\frac{\varepsilon_{12}(0)}{T \xi_c} \right)^d
	\theta \left(\xi_c-\frac{\varepsilon_{12}(0)}{T} \right).
\label{probdens}
\end{eqnarray}
Here $v_d=2 \pi^{d/2} /\Gamma (d/2)d$ is the volume of a $d$-dimensional
sphere of unit radius.
For the VRH problem, the energies of the two sites are not fixed but rather
are governed by distribution functions. Therefore the normalized distribution
of energies $f_{1}(\varepsilon_2)$ of sites which can be connected to a site
in the system is
\begin{eqnarray}
\Upsilon_{1}f_1(\varepsilon_2)
        &=&\int d\varepsilon_1 G_{AA}(\varepsilon_1,\varepsilon_2)
	f_0(\varepsilon_1)  \nonumber \\
	&=&g_A v_d \left(\frac{a\xi_c}{2} \right)^d
	\int \left(1-\frac{\varepsilon_{12}(0)}
	{T\xi_c} \right)^d \theta \left( 1-\frac{\varepsilon_{12}(0)}
{T\xi_c} \right)
	f_0(\varepsilon_1) d\varepsilon_1.
\label{Nf}
\end{eqnarray}
Here $f_0(\epsilon_1)$ is the distribution of energies of the first
site; the normalization coefficient $\Upsilon_1$ is the
average number of sites which can be directly linked to the first one
(i.e., the average number of bonds per site). Repeating the procedure
of Eq.~(\ref{Nf}) several times, we find that the distributions of
site energies obtained in this manner converge rapidly to the
eigenfunction of the dimensionless integral equation
\begin{equation}
\lambda_d f(\Delta_2)=\int[1-\Delta_{12}(0)]^d \theta(1-\Delta_{12}(0))
f(\Delta_1) d \Delta_1.
\label{int2}
\end{equation}
We conjecture that this eigenfunction is approximately equal to the
distribution of energies for the sites in the infinite cluster.  The
eigenvalue $\lambda_d$ is proportional to $\Upsilon$ which would then be
the average number of bonds per site in the infinite cluster,
\begin{equation}
\Upsilon=\lambda_d g_A v_d \left( \frac{a \xi_c}{2} \right)^d T \xi_c.
\label{A}
\end{equation}
We further conjecture that
$\Upsilon$ is an invariant of the percolation problem, and equals the
average number of bonds per site $B_c$ for the random sites (RS)
problem\cite{ShklEfr}. With this assumption, we obtain
\begin{equation}
\lambda_d=2\frac{n_c^{\rm RS}}{n_c^{\rm VRH}},
\label{lambda}
\end{equation}
where $n_c^{\rm RS}$ and $n_c^{\rm VRH}$ are the critical concentrations of
the  RS and VRH percolation problems.

A numerical solution of the integral equation (\ref{int2}) gives:
$\lambda_2=0.4301,\lambda_3=0.3154, \lambda_4=0.2489$,  in excellent
agreement with the values given by Eq.~(\ref{lambda}), with
$n_c^{\rm RS}$ and $n_c^{\rm VRH}$ obtained via simulations of the
corresponding  percolation problems: $\lambda_2=0.410 \pm 0.004,
\lambda_3=0.303 \pm 0.004, \lambda_4=0.232 \pm 0.005$.  The simulation
procedures we used to find $n_c^{\rm RS}$ and $n_c^{\rm VRH}$ are nearly
identical to the one described in Section~\ref{simulation} and were run on
$N=8100$ sites and averaged over 50 runs  for $d=3$ and $d=4$ and on
$N=6400$ and averaged over 100  runs for
$d=2$.

\subsection{Generalization to the ``Colored'' Percolation Problem}
\label{AB_invariant}

We shall now generalize the discussion of the previous section to the
case of a two-color model, $g_B\neq0$. The sites are now characterized
by two parameters: a continuous variable $\epsilon$ and a discrete
variable, $A$ or $B$. Consequently, instead of a single
equation (\ref{int2}), we now have a system of two integral equations:
\begin{equation}
    \Upsilon
        \left(
            \begin{array}{l}
            f_A(\varepsilon_2)\\
            f_B(\varepsilon_2)
            \end{array}
        \right)
   = \int\left(
            \begin{array}{ll}
    G_{AA}(\varepsilon_1,\varepsilon_2) &
G_{AB}(\varepsilon_1,\varepsilon_2)\\
 \gamma^{-1}  G_{AB}(\varepsilon_2,\varepsilon_1) &
		\gamma^{-1}  G_{AA}(\varepsilon_1,\varepsilon_2)
            \end{array}
        \right)
        \left(
            \begin{array}{l}
            f_A(\varepsilon_1)\\
            f_B(\varepsilon_1)
            \end{array}
        \right) d \varepsilon_1.
\label{asymmetric}
\end{equation}
Here, $G_{AA}(\varepsilon_1,\varepsilon_2)$ is given by
Eq.~(\ref{probdens}).  The off-diagonal element
$G_{AB}(\varepsilon_1,\varepsilon_2)$ is the probability density of
finding a link between an $A$-state with energy $\varepsilon_1$ and a
$B$-state with energy $\varepsilon_2$. It can be obtained from the
probability density $G_{AA}$ given by Eq.~(\ref{probdens}) by making
the replacement $\varepsilon_{12}(0) \rightarrow
\varepsilon_{12}(H)$.

The kernel of this integral equation is not symmetric.
However it can be symmetrized by performing a linear transformation:
\begin{equation}
        \left(
            \begin{array}{l}
            f_A\\
            f_B
            \end{array}
        \right)
     =
        \left(
            \begin{array}{ll}
            1& 0\\
            0 & \gamma^{-1/2}
            \end{array}
        \right)
        \left(
            \begin{array}{ll}
            h_{A}\\
            h_{B}
            \end{array}
        \right).
\label{weight}
\end{equation}
Clearly, this transformation preserves the eigenvalues.
In the new basis, the integral equation (\ref{asymmetric}) attains a
symmetric form
\begin{equation}
    \Upsilon
        \left(
            \begin{array}{l}
            h_A(\varepsilon_2)\\
            h_B(\varepsilon_2)
            \end{array}
        \right)
   = \int\left(
            \begin{array}{ll}
      G_{AA}(\varepsilon_1,\varepsilon_2) &
\gamma^{-1/2} G_{AB}(\varepsilon_1,\varepsilon_2)\\
      \gamma^{-1/2} G_{AB}(\varepsilon_2,\varepsilon_1) &
 \gamma^{-1}G_{AA}(\varepsilon_1,\varepsilon_2)
            \end{array}
        \right)
        \left(
            \begin{array}{l}
            h_A(\varepsilon_1)\\
            h_B(\varepsilon_1)
            \end{array}
        \right) d \varepsilon_1,
\label{symmetric}
\end{equation}
guaranteeing that all of its eigenvalues are real.

We now extend the method of invariance to the ``two-color''
percolation problem and conjecture that the maximal eigenvalue
$\Upsilon$ is independent of the magnetic field. Let us show that by
assuming the invariance of $\Upsilon$ we recover the results of the
weak field perturbative approach described in Section~\ref{weak}.
First we solve equation (\ref{symmetric}) at $H=0$, for which
$G_{AB}(\varepsilon_1,\varepsilon_2)=G_{AA}(\varepsilon_1,\varepsilon_2)$.
One can easily see that the normalized solution corresponding to the
maximal eigenvalue has the form
\begin{equation}
  |0\rangle\equiv
        \left(
            \begin{array}{l}
            h_A(\varepsilon_1)\\
            h_B(\varepsilon_1)
            \end{array}
        \right)
     =
        \frac{1}{\sqrt{1+\gamma}}
        \left(
            \begin{array}{l}
            \gamma^{1/2}\\
            1
            \end{array}
        \right)f(\varepsilon_1).
\label{solution}
\end{equation}
Here the function $f(\varepsilon)$ is defined as the normalized solution of
the ``monochromatic'' integral equation
\begin{equation}
\Lambda_0 f(\varepsilon_2) = \int G_{AA}(\varepsilon_1,\varepsilon_2)
f(\varepsilon_1) d \varepsilon_1,
\end{equation}
corresponding to the maximal possible eigenvalue $\Lambda_0$.

As we apply a small magnetic field $H$, the percolation threshold
$\xi_c$ shifts away from its zero field value, $\xi_c^0$.  However,
according to our conjecture, the eigenvalue $\Upsilon$ remains
constant.  To first order in $\mu_BH$ and $\delta\xi_c$, the
correction to $\Upsilon$ is simply $\delta\Upsilon=\langle 0| \delta G|0
\rangle$, in complete analogy with first order perturbation
theory in quantum mechanics.  Thus the field induced correction to
$\Upsilon$ is given by the following relation:
\begin{equation}
\delta  \Upsilon
   =\frac{1}{1+\gamma}
    \int f(\varepsilon_2)(\gamma^{1/2},1)
        \left(
            \begin{array}{ll}
            \delta G_{AA}(\varepsilon_1,\varepsilon_2) & \gamma^{-1/2}
		\delta G_{AB}(\varepsilon_1,\varepsilon_2)\\ \gamma^{-1/2}
\delta G_{AB}(\varepsilon_2,\varepsilon_1) & \gamma^{-1}
		\delta G_{AA}(\varepsilon_1,\varepsilon_2)
            \end{array}
        \right)\!
        \left(
            \begin{array}{l}
            \gamma^{1/2}\\
            1
            \end{array}\!
        \right) \!
     f(\varepsilon_1)d\varepsilon_1 d\varepsilon_2.
\label{deltalambda}
\end{equation}
The condition $\delta \Upsilon = 0$ may be rewritten in the form
\begin{equation}
    \int f(\varepsilon_2)
        [(1+\gamma^2)\delta G_{AA}(\varepsilon_1,\varepsilon_2) +
	\gamma(\delta G_{AB}(\varepsilon_1,\varepsilon_2)
         +\delta G_{AB}(\varepsilon_2,\varepsilon_1))]
     f(\varepsilon_1)d \varepsilon_1 d \varepsilon_2 = 0.
\label{condition}
\end{equation}
This equation enables one to find the magnetic field induced correction to
$\xi_c^0$, since
\begin{eqnarray}
G_{AA}(\varepsilon_1,\varepsilon_2)
&=&G\left(\xi_c-\frac{\varepsilon_{12}(0)}{T}\right),\\
G_{AB}(\varepsilon_1,\varepsilon_2)
&=&G\left(\xi_c-\frac{\varepsilon_{12}(H)}{T}\right).
\end{eqnarray}
To first order in $\delta \xi_c$ and $\mu_B H$, the condition
$\delta\Upsilon =0$ (\ref{condition})  may be rewritten as
\begin{equation}
    \int f(\varepsilon_2)f(\varepsilon_1)
	G'\left(\xi_c^0-\frac{\varepsilon_{12}(0)}{T}\right)
        \left\{\delta \xi_c (1+\gamma)^2  -
\gamma\frac{H}{T}\left.\frac{\partial(\varepsilon_{12}+
{\varepsilon_{21}})}{\partial H}\right|_{H=0}
     \right\}d \varepsilon_1 d\varepsilon_2 = 0.
\label{condition2}
\end{equation}
{}From the definition (\ref{var12}) of $\varepsilon_{12}(H)$, one finds
\begin{equation}
\frac{\partial}{\partial H} \left[\varepsilon_{12}(H)  +
\varepsilon_{21}(H) \right]=
\mu_B \left[1+{\rm sign}(\varepsilon_2-\varepsilon_1)\right]+
\mu_B \left[ 1+{\rm sign}(\varepsilon_1-\varepsilon_2) \right]
=2\mu_B.
\label{simple}
\end{equation}
Using Eqs. (\ref{condition2}) and (\ref{simple}), we recover the result
of the perturbative approach:
\begin{equation}
\delta \xi_c = \frac{\gamma}{(1+\gamma)^2}\frac{2\mu_B H}{T}.
\end{equation}

To consider the case of arbitrarily strong magnetic fields, we change to the
dimensionless variables $\Delta_1$ and $\Delta_2$ and, in doing so, reduce the
integral eigenvalue equation to a dimensionless form which can
then be solved numerically:
\[
    \lambda_d \left(
            \begin{array}{l}
            h_A(\Delta_2)\\
            h_B(\Delta_2)
            \end{array}
        \right)=
\]
\begin{equation}
   \int\!\left(\!
            \begin{array}{ll}
            \left(1 - \Delta_{12}(0)\right)^d
      \theta\left(1 - \Delta_{12}(0)\right) & \gamma^{-1/2}
\left(1 - \Delta_{12}(\chi)\right)^d
      \theta\left(1 - \Delta_{12}(\chi)\right)\\
            \gamma^{-1/2}\left(1 - \Delta_{21}(\chi)\right)^d
      \theta\left(1 - \Delta_{21}(\chi)\right) & \gamma^{-1}
\left(1 - \Delta_{21}(0)\right)^d
      \theta\left(1 - \Delta_{21}(0)\right)
            \end{array}
        \!\right)
        \left(\!
            \begin{array}{l}
            h_A(\Delta_1)\\
            h_B(\Delta_1)
            \end{array}
        \!\right) d\Delta_1.
\label{dimensionless}
\end{equation}
Here, the eigenvalue $\lambda_d$ is related to $\Upsilon$ by
Eq.~(\ref{A}).  For the two-color problem,
$\lambda_d=\lambda_d(\chi,\gamma)$, and the invariance conjecture is
equivalent to the assumption that the product $\lambda_d(\chi,\gamma)
[\xi_c(\chi,\gamma)]^{d+1}$ is independent of both the magnetic field
and the ratio $\gamma$ of the two densities of states.  Therefore
using Eq.~(\ref{A}), the universal scaling function
$F(x)\equiv\xi_c(T,H)/\xi_c^0$ can be expressed in terms of the
eigenvalue $\lambda_d$,
\begin{equation}
F=\left[\frac{\lambda_d(0,0)}
{\lambda_d(\chi,\gamma)}\right]^{1/(d+1)}.
\label{*!*}
\end{equation}
Formula (\ref{*!*}) expresses $F$ in terms of $\chi=\mu_BH/T\xi_c$,
instead of $x=\mu_BH/T\xi_c^0$. Noting that
$x/\chi=\xi_c/\xi_c^0\equiv F$, we find:
\begin{equation}
x=\chi\left[\frac{\lambda_d(0,0)}
{\lambda_d(\chi,\gamma)}\right]^{1/(d+1)}.
\label{*!!*}
\end{equation}
Eqs.~(\ref{*!*}) and (\ref{*!!*}) determine $F(x)$ in parametric form.

In order to assess the validity of the invariant approach, we solved
Eq.~(\ref{dimensionless}) numerically at the two values of $\gamma$
used in Section \ref{simulation}, $\gamma=1$ and $1/2$, over the
entire relevant range of dimensionless magnetic fields.  The resulting
plots of $F(x)$ are shown in Fig 2ab together with the two plots
obtained via simulations.  Let us first examine the two curves
obtained for $\gamma=1/2$ (Fig 2a).  It is clear that at weak fields
the invariant method yields an accurate approximation of the universal
scaling function $F(x)$ at $\gamma=1/2$.  In addition, we see that at
$\gamma=1/2$ this approach gives the proper limiting value of $F(x)$.
At intermediate fields, the function $F(x)$ obtained from the
invariant method continues smoothly between its limiting forms,
deviating from the corresponding values of $F(x)$ obtained from
simulations by at most $2\%$.

Upon examination of the two curves in Fig 2b, it is equally clear that at
$\gamma=1$ the invariant method accurately reproduces the scaling function
$F(x)$ at small fields.  This method gives the proper limiting value
of $F(x)$ at $\gamma=1$ which, however, is higher than the corresponding
simulation value by approximately $1.5\%$.  This disagreement can be
attributed to the systematic suppression of $F(x)$ at large $x$ due to certain
finite size effects in the simulations run at $\gamma=1$
(see Section\ref{simulation}).  The function $F(x)$ obtained from the
invariant method smoothly continues between its limiting forms, differing
from the corresponding values of $F(x)$ obtained from simulations by at most
$6\%$.  We believe that the magnitude of this discrepancy between the scaling
functions, determined by simulations and the invariant method, at
intermediate fields is considerably enhanced by the above mentioned finite
size  effects in simulations run at $\gamma=1$.

We should point out that, for the values of $F(x)$ obtained from the
simulations, the standard error in $F(x)$ is approximately $0.005$.
Thus at $\gamma=1/2$ there is a wide range of intermediate fields
within which the difference between the numerically and analytically
determined scaling functions is larger than the standard error.  For
$\gamma=1$ the two curves differ by more than the standard error for
$x \geq 0.05$.  Thus the invariant method does not suffice to
accurately determine the universal scaling function $F(x)$.  We see
two possible reasons for this shortcoming: the first being that this
procedure might not yield information about the infinite cluster and
the second being the possibility that the average number of bonds per
site, for sites belonging to the infinite cluster, is not an invariant
of the dimensionless percolation problem.  These questions are the
subject of future work.

\section{Conclusion}

In this paper, we considered the problem of hopping magnetoresistance
induced by Zeeman splitting. The problem has been reduced to the
calculation of a universal function $F(x)$ of a single dimensionless
parameter $x$ (which depends upon magnetic field and
temperature).  To find $F(x)$, one has to solve certain dimensionless
``two-color'' percolation problem.  We found the limiting behavior of
$F(x)$ analytically and obtained its full shape numerically for two
values of the ratio $g_A/g_B=1$ and $g_A/g_B=1/2$.  In addition, we
developed an approximate method which enables us to solve the ``two
color'' percolation problem analytically at any $x$.
This approach gives a satisfactory extrapolation of the function
$F(x)$ between its two limiting forms.

This theory can be applied to any system of localized electrons for
which the width of the distribution function of the energies of the
localized states is larger than the on-site Coulomb repulsion energy.
Bulk amorphous silicon satisfies this condition\cite{Ephron,Ephron2}.
In the bulk ($d=3$), there exists a range of temperatures
(\ref{restriction}) within which Zeeman splitting makes the dominant
contribution to the magnetoresistance.  As was proposed
in\cite{Ephron2}, this theory can then be used to probe the relative
concentration of singly and doubly occupied sites which contribute to
transport.

This work was initiated by extensive communications with M. R. Beasley
and D. Ephron.  We are also indebted to A. S. Ioselevich and B. I. Shklovskii
for helpful discussions. This work was supported by NSF Grant DMR-9423244.
One of us (PC) acknowledges the Department of Education for fellowship
support.

\narrowtext

\begin{figure}
\caption{The universal function, $F(x)$ for $d=2$. The procedure for
extracting $F(x)$ and $x$ from the simulation data gives the standard
errors 0.005 for $F(x)$ and less than 0.003 for $x$. Insets: normalized
critical site concentration as a function of dimensionless magnetic field
$\chi$.  The standard error of $n_c(\chi)/n_c(0)$ is 0.01 on average.
(a): $\gamma=1$; for $x \ge x_{\rm th}$, there is a systematic suppression of
$F(x)$ below its true value, $F(x_{\rm th})=1.26$; Inset: a linear fit to the
data  for $\chi \le 0.20$ gave a slope of $1.47 \pm 0.03$ in good agreement
with  the theoretically predicted value, $3/2$.  (b):$\gamma=1/2$; for $x \ge
x_{\rm th}$,  the function $F(x)$ attains its exact value, $F(x_{\rm
th})=1.14$; Inset: a linear fit  to the data taken at $\chi \le 0.2$ yielded
a slope of $1.33 \pm 0.04$ in good  agreement with the value $ 4/3$ following
from the perturbation approach.}
\label{fig1}
\end{figure}

\begin{figure}
\caption{The approximate scaling function $F(x)$ obtained via the invariant
approach, for $d=2$, shown with the corresponding curves extracted from
simulations.  (a): $\gamma=1$; the analytically derived scaling function
reproduces  the simulation results only at very small values of x.  The
difference in the  limiting values of the two curves can be attributed to the
finite size effect,  described earlier, which is present in simulations run at
$\gamma=1$.   In addition , this finite size effect broadens the range of the
scaling variable $x$, within which the two curves differ substantially. (b):
$\gamma=1/2$; for $x\ll x_{\rm th}$ and $x\geq x_{\rm th}$, the  function
$F(x)$ obtained via the invariant approach is in  good agreement with the
numerically determined function.  Within a large range  of intermediate
values of the scaling variable $x$, however, the two functions  differ by
more than 0.005, the standard error of the numerically determined  $F(x)$.}
\label{fig2}
\end{figure}


\begin{references}
\bibitem{singlet} Recently this assumption has been extensively
tested, see Ref.\cite{Ephron2}.
\end{references}
\end{document}